\documentclass[sigconf]{acmart} 

\usepackage{booktabs}
\usepackage{multirow}
\usepackage{array}      
\usepackage{tabularx} 
\usepackage{wrapfig}
\usepackage{dblfloatfix}
\usepackage{graphicx}
\usepackage{makecell}
\usepackage{balance}
\usepackage{enumitem}
\usepackage{hyperref}
\usepackage{pifont}
\usepackage[ruled,vlined]{algorithm2e}

\copyrightyear{2026}
\acmYear{2026}
\setcopyright{cc}
\setcctype{by}                      
\acmConference[KDD '26]{Proceedings of the 32nd ACM SIGKDD Conference on Knowledge Discovery and Data Mining V.2}{August 9-13, 2026}{Jeju Island, Republic of Korea.}
\acmBooktitle{Proceedings of the 32nd ACM SIGKDD Conference on Knowledge Discovery and Data Mining V.2 (KDD 2026), August 9-13, 2026, Jeju Island, Republic of Korea}
\acmISBN{979-8-4007-2259-2/2026/08}
\acmDOI{10.1145/3770855.3818987}
\author{Yongqing Jiang}
\affiliation{%
  \institution{Sichuan University}
  \city{Chengdu}
  \country{China}
}
\email{yongqingjiang@stu.scu.edu.cn}

\author{Jianze Wang}
\affiliation{%
  \institution{Sichuan University}
  \city{Chengdu}
  \country{China}
}
\email{jzwang@scu.edu.cn}

\author{Zhiqi Shen}
\affiliation{%
  \institution{Nanyang Technological University}
  \city{Singapore}
  \country{Singapore}
}
\email{zqshen@ntu.edu.sg}

\author{Zhenghong Lin}
\affiliation{%
  \institution{Nanyang Technological University}
  \city{Singapore}
  \country{Singapore}
}
\email{hongzhenglin970323@gmail.com}

\author{Jiayuan Wang}
\affiliation{%
  \institution{Fuzhou University}
  \city{Fuzhou}
  \country{China}
}
\email{wwwangjiayuan@gmail.com}

\author{Yijian Yang}
\affiliation{%
  \institution{Sichuan University}
  \city{Chengdu}
  \country{China}
}
\email{yijianyang@alu.scu.edu.cn}

\author{Kaoshan Dai\texorpdfstring{\textsuperscript{*}}{*}}
\affiliation{%
  \institution{Sichuan University}
  \city{Chengdu}
  \country{China}
}
\email{kdai@scu.edu.cn}

\author{Haoran Luo\texorpdfstring{\textsuperscript{*}}{*}}
\affiliation{%
  \institution{Nanyang Technological University}
  \city{Singapore}
  \country{Singapore}
}
\email{haoran.luo@ieee.org}

\renewcommand{\shortauthors}{Jiang et al.}  

\begin{document}

\title[Toward Physically Consistent and Simulation-Executable Programmatic Generation]{Rethinking Scientific Modeling: Toward Physically Consistent and Simulation-Executable Programmatic Generation}

\begin{abstract}
Structural modeling is a fundamental component of computational engineering science, in which even minor physical inconsistencies or specification violations may invalidate downstream simulations. The potential of large language models (LLMs) for automatic generation of modeling code has been demonstrated. However, non-executable or physically inconsistent outputs remain prevalent under stringent engineering constraints. A framework for physics-consistent automatic building modeling is therefore proposed, integrating domain knowledge construction, constraint-oriented model alignment, and verification-driven evaluation. CivilInstruct is introduced as a domain-specific dataset that formalizes structural engineering knowledge and constraint reasoning to enable simulation-ready model generation. A two-stage fine-tuning strategy is further employed to enforce constraint satisfaction and application programming interface compliance, substantially reducing hallucinated and non-conforming outputs. MBEval is presented as a verification-driven benchmark that evaluates executability and structural dynamics consistency through closed-loop validation. Experimental results show consistent improvements over baselines across rigorous verification metrics. Our code is available at \url{https://github.com/Jovanqing/AutoBM}.
\end{abstract}



\maketitle
\begingroup
  \renewcommand{\thefootnote}{\fnsymbol{footnote}}
  \footnotetext[1]{Corresponding authors}
\endgroup

\begin{figure}[!htb]
  \centering
  \includegraphics[width=0.98\linewidth]{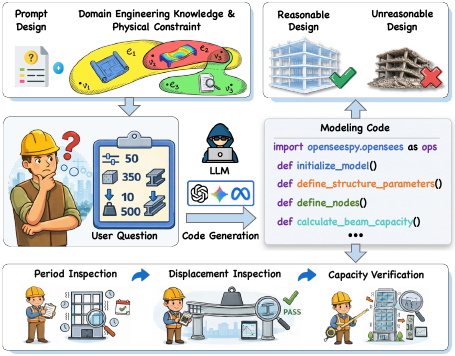}
  \caption{Task formulation of llm-driven automatic building modeling from natural language descriptions}
  \label{fig:matrix}
\end{figure}

\begin{figure*}[t]
  \centering
  \includegraphics[width=\textwidth]{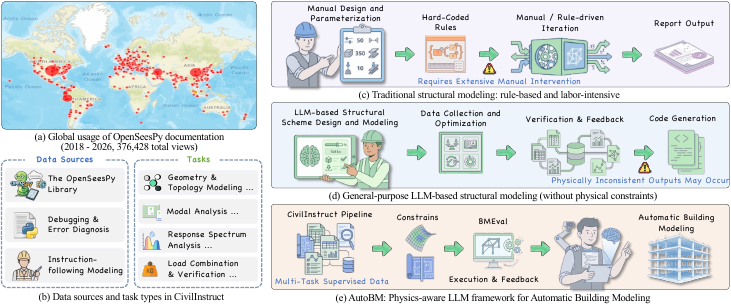} 
  \caption{Overview of this study, illustrating the integration of large language models with structural engineering knowledge for Automatic Building Modeling (AutoBM), including task motivation, data foundations, and the framework}
  \label{fig:study_overview}
\end{figure*}

\section{Introduction}

Accurate numerical modeling is a cornerstone of computational structural science. Recent advances in large language models (LLMs) \cite{NIPS2017_3f5ee243, NEURIPS2022_9d560961, luo2025kbqa, adcock2026llama} introduce a new paradigm for scientific modeling. As shown in Figure \ref{fig:matrix}, LLMs can interpret natural-language specifications and generate corresponding structural modeling code \cite{seed2025seed, guo2024deepseek, huang2025opencoder}, enabling automated construction of simulation-ready models. Within this emerging paradigm, structural modeling stands out as a compelling domain due to its complexity, rule constraints, and reliance on expert knowledge. OpenSeesPy is a cornerstone platform for seismic and structural analysis \cite{yan2025opstool, batukan2024modularbuildingpy, sdeghzadeh2025modeling, zhu2018openseespy} due to its strong capability in simulating complex structural systems (Figure \ref{fig:study_overview}(a)). When integrated with OpenSeesPy and multi-source data (Fig. \ref{fig:study_overview}(b)), LLMs enable automatic building modeling (AutoBM) across multiple building types by directly generating modeling code. Relative to rule-based template methods (Fig. \ref{fig:study_overview}(c)), LLM-driven modeling offers superior scalability, positioning it as a promising computational interface for engineering simulation.

Despite this promise, existing LLM-based approaches to structural modeling remain fundamentally limited. Automated approaches based on LLMs are increasingly applied to modeling research (e.g., structural verification \cite{liang2025automating}, 2D frame modeling \cite{geng2025lightweight}, integration with the computational engine \cite{liang2025integrating} of OpenSeesPy). These LLM-driven modeling approaches \cite{bedagkar2025llm, bougie2025citysim, duan2025llm} have been shown to improve the efficiency of building modeling. However, prior work largely overlooks task-specific fine-tuning for structural engineering, despite the domain’s strict requirements on engineering compliance beyond syntactic correctness. As shown in Figure \ref{fig:study_overview}(d), without domain-specific knowledge integration, ensuring the model's physical correctness is difficult for general-purpose LLMs.

This raises a key question: \textbf{\textit{"How can LLMs be systematically constrained to enable structurally and physically valid programmatic generation for building modeling?"}} To address this question, we propose a physics-aware LLM framework for AutoBM, as shown in Figure \ref{fig:study_overview}(e). The main contributions of this work can be summarized as follows: 

\begin{itemize}[leftmargin=*, itemsep=0.3em]
\item \textbf{Novel AutoBM Task:} We formalize Automatic Building Modeling (AutoBM) as a research task with clearly defined problem boundaries. The task specifies user-driven modeling requirements by defining task inputs (building function, plan dimensions) and task outputs (engineering-compliant, high-precision structural modeling code).

\item \textbf{BMBench Dataset:} We propose BMBench, a comprehensive AutoBM benchmark integrating large-scale instruction data and multidimensional evaluation metrics across 10,912 samples and 128 cases.

\item \textbf{RLA-SPC:} We propose a reinforcement learning alignment strategy under structural physical constraints, and experimental results show that RLA-SPC improves LLM compliance with engineering physical constraints.

\end{itemize}

\section{Related Work}
\textbf{Large Language Models.} General-purpose LLMs \cite{achiam2023gpt, comanici2025gemini, yang2025qwen3} exhibit limited generalization in specialized domains, prompting domain adaptation efforts and the development of domain-specific benchmarks across medicine \cite{singhal2025toward}, finance \cite{wu2023bloomberggpt}, and biology \cite{luo2022biogpt}. Within the civil engineering domain, applied research on LLMs exhibits a trend toward multifaceted breakthroughs. For example, Kim et al. \cite{kim2025can} and Pandey et al. \cite{pandey2025openfoamgpt} demonstrated the feasibility of using LLMs to generate high-fidelity simulation codes for fluid-structure interaction and computational fluid dynamics (CFD), respectively. Qin et al. \cite{qin2024intelligent} developed an intelligent shear wall design framework that achieved substantial gains in design efficiency while satisfying mechanical constraints. Jiang et al. \cite{jiang2025large, jiang2026multitask} and Jeoung et al. \cite{jeoung2025zero} applied visual-language modeling techniques to automate visual analytics tasks, including post-earthquake damage assessment and construction site monitoring.

\begin{figure*}[t]
  \centering
  \includegraphics[width=\textwidth]{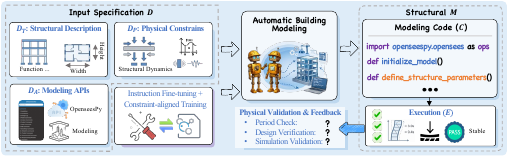} 
  \caption{The definition and overview of the AutoBM task.}
  \label{fig:definition and overview}
\end{figure*}

\textbf{Automatic Structural Modeling and Reasoning.} Early computational approaches for design–analysis mapping mainly relied on rigid scripting interfaces \cite{zareian2010evaluation, heshmati2021fema, lignos2010steel} or parametric template-based generation methods \cite{guan2020python, nuzzo2020dibrast, zhou2023automated, beutler2022automated}. With the emergence of LLMs, existing studies have explored domain adaptation through model fine-tuning \cite{xue2024question} and the integration of structured domain knowledge (e.g., knowledge graphs \cite{yang2024prompt, li2024intelligent}, material mechanism discovery \cite{guo2024knowledge}, energy system optimization \cite{gonccalves2025empirically}). More recent research has further advanced toward system-level reasoning by adopting multi-agent collaboration mechanisms \cite{ye2025mas, chen2023multi, kretzschmar2025using}, while the incorporation of chain-of-thought reasoning and literate programming paradigms has enabled LLMs to orchestrate end-to-end simulation pipelines \cite{zhang2024renaissance}, motivating the development of domain-specific benchmarks for systematic evaluation of LLM-driven engineering automation \cite{avila2025toward, li2025drafterbench, wan2025som, ersoz2024aidcon}.

\section{Task: AutoBM}
The AutoBM task is formally defined as the process of generating executable, physically consistent structural modeling code from structured engineering specifications using LLMs. As shown in Figure \ref{fig:definition and overview}, the fundamental objective of AutoBM is to bridge the gap between natural language intent and rigorous structural engineering requirements while strictly adhering to physical constraints.
Formally, a building modeling specification is defined as $\mathcal{S} = (D_T, D_P, D_A)$. In this formulation, the textual description of the structural modeling task, including building geometry, building function, and seismic intensity, is denoted by $D_T$. The set of physical and engineering constraints (e.g., structural mechanics principles, design verification requirements) is represented by $D_P$. Furthermore, the target modeling APIs and the computational environment (e.g., OpenSeesPy functions, modeling workflows) are specified by $D_A$. Given this specification, the modeling target is characterized as a structural model implementation $\mathcal{M}=(C, E)$, where $C$ represents the generated modeling code and $E$ represents the execution outcomes. These outcomes encompass successful compilation, runtime stability, and physically valid responses (e.g., modal periods, internal forces).

The AutoBM task is characterized as the generation of an optimal structural model $\hat{M}$, conditioned upon the input specification $D$. This optimization objective is formulated as:

\begin{equation}
\hat{M} = \arg\max_{M} \Pr(M \mid D)
\end{equation}

In contrast to general-purpose code-generation paradigms, the AutoBM task requires the generated code to satisfy constraints on both syntactic correctness and physical feasibility. Consequently, the optimization objective is characterized as multi-dimensional. A composite objective function $\vec{F}(M)$ is formulated as follows:

\begin{equation}
\resizebox{0.95\columnwidth}{!}{$
\max \vec{F}(M)
= \alpha_1 S_{\mathrm{Exec}}(M \mid D_A)
+ \alpha_2 S_{\mathrm{Phys}}(M \mid D_P)
+ \alpha_3 S_{\mathrm{Spec}}(M \mid D_T)
$}
\end{equation}


Here, $S_{\text{Exec}}(M \mid D_A)$ evaluates executability and API compliance, while $S_{\text{Phys}}(M \mid D_P)$ measures physical consistency and structural validity under mechanics principles and engineering constraints. Semantic alignment with modeling requirements is assessed by $S_{\text{Spec}}(M \mid D_T)$. The trade-offs among these objectives are regulated by non-negative weights $\alpha_i$, balancing code flexibility and physical correctness. The objective of AutoBM is therefore to generate executable, engineering-valid structural models, thereby forming a closed-loop paradigm for LLM-based building modeling.

\begin{figure*}[t]
  \centering
  \includegraphics[width=\textwidth]{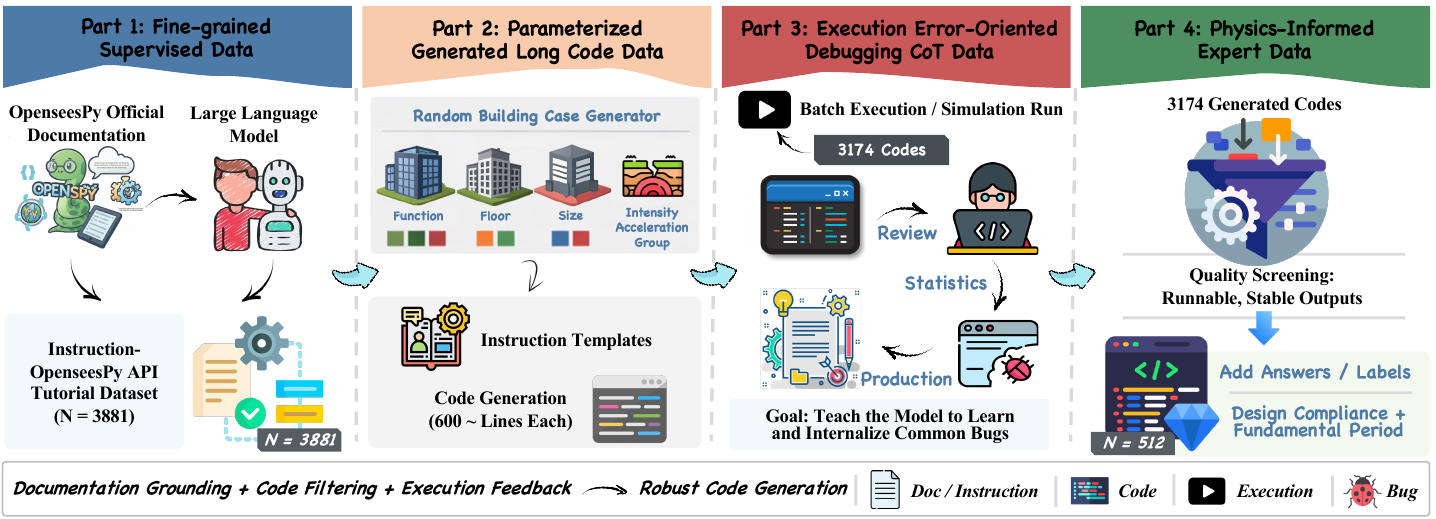} 
  \caption{Overview of the CivilInstruct construction procedure.}
  \label{fig:CivilInstruct construction}
\end{figure*}

\section{BenchMark: BMBench}
This section introduces the modeling of building benchmark (BMBench), a novel benchmark designed for the training and evaluation of AutoBM task. The construction process, evaluation protocol, and the metrics derived from this protocol are described in detail.

\subsection{CivilInstruct Dataset Construction}
The construction of the CivilInstruct dataset follows a four-stage pipeline, as shown in Figure \ref{fig:CivilInstruct construction}. First, OpenSeesPy documentation was combined with LLM-based synthesis to generate 3,800 API-learning samples and 3,100 expert-level instruction-code instances covering building functions and seismic conditions. Subsequently, 3,500 Bug-CoT samples were derived from execution error logs, and 512 high-quality code instances were filtered and annotated with structural periods obtained via finite element analysis, providing physically grounded supervisory signals. Further implementation details for each stage are provided in Appendix \ref{sec:bmbench_impl}.

\subsection{BMEval: Benchmark Design}
To ensure the quality, relevance, and effectiveness of the AutoBM task, BMEval was constructed through a three-stage process. First, building samples were randomly generated to cover a diverse set of attributes (e.g., building function, floor height, plan dimensions) within a reasonable structural parameter space. Then, an expert-driven mechanism was adopted to optimize and validate all design solutions in strict accordance with seismic design codes, ensuring their structural feasibility and executability. Subsequently, empirical formulas were used to compute the fundamental period of each structure as ground-truth reference values. In total, 128 labeled evaluation samples were obtained, each containing complete building information and its corresponding structural period.

To evaluate LLM performance on the AutoBM task, we adopt a multidimensional evaluation protocol covering program executability, computational accuracy, and engineering compliance. For each test sample, the model independently generates $n$ candidate programs, which are executed sequentially in a restricted sandbox. A program is deemed successful if it terminates normally within the prescribed time limit. Following the standard evaluation paradigm of mainstream code-generation benchmarks (e.g., HumanEval \cite{raihan2025mhumaneval}, MBPP \cite{yu2025humaneval}), we employ the unbiased $Pass@k$ estimator to measure the probability that at least one valid solution is obtained:

\begin{equation}
\text{pass@k} =
\begin{cases}
1, & n \le k, \\
1 - \frac{\binom{n-c}{k}}{\binom{n}{k}}, & \text{otherwise}.
\end{cases}
\end{equation}

Building upon $Pass@k$, we further define three task-specific metrics: $Pass@k_{\text{period}}$, $Pass@k_{\text{compliance}}$, and $Pass@k_{\text{strict}}$, corresponding to structural cycle consistency, engineering compliance, and their joint assessment. Formal definitions of these metrics are provided in Appendix \ref{sec:BMEval Construction Details}. See Appendix \ref{sec:Prompt used in AutoBM} for the prompt used to construct the BMBench.

\section{Methodology: Two-Stage Reinforcement Learning Alignment Under Structural Physical Constraints}

We formulate AutoBM as a conditional program synthesis problem. Given an engineering instruction $x$ (e.g., building function, plan dimensions), a language model parameterized by $\theta$ defines a conditional policy $\pi_\theta(\cdot\mid x)$ over OpenSeesPy code sequences $y = (y_1,\ldots,y_T)$. Our goal is to learn $\pi_\theta$ that produces code that is (i) domain-compliant, (ii) logically complete, and
(iii) physically valid under structural mechanics constraints.
To this end, we propose RLA-SPC, a two-stage reinforcement learning framework for aligned policy optimization. As shown in Algorithm~\ref{alg:rla-spc}, the policy is first initialized via supervised fine-tuning to ensure basic executability, and subsequently optimized using SPC-GRPO to enforce structured and physics-aware constraints.

\subsection{Stage I: Domain Instruction Fine-Tuning}
Let $\mathcal{D}=\{(x^{(n)},y^{(n)})\}_{n=1}^{N}$ denote the instruction dataset (CivilInstruct),
where $x$ is an AutoBM instruction and $y$ is a reference OpenSeesPy implementation.
We fine-tune a pretrained model by maximizing the conditional likelihood of the reference code:

\begin{equation}
\mathcal{L}_{\mathrm{SFT}}(\theta)
= -\mathbb{E}_{(x,y)\sim\mathcal{D}}
\left[\sum_{t=1}^{|y|}
\log \pi_{\theta}(y_t \mid x, y_{<t})\right].
\label{eq:sft}
\end{equation}

The policy is denoted as $\pi_{\mathrm{ref}}$ (the \emph{reference policy}), which captures domain syntax, OpenSeesPy API usage patterns, and engineering priors. However, SFT alone does not guarantee physical validity because it does not explicitly optimize for physics-based criteria.

\subsection{Stage II: Structural Physics-Constrained Reinforcement Learning Alignment}

To explicitly enforce physical consistency, we further align the model using a physics-constrained variant of group-relative policy optimization (SPC-GRPO). SPC-GRPO follows the group-relative optimization paradigm, while incorporating domain-specific physical constraints via a multi-granularity hybrid reward and a reference-policy KL regularization. For each query $q$ (we use $q \equiv x$), we sample a group of $G$ candidate outputs $\{o_i\}_{i=1}^{G} \sim \pi_{\theta_{\mathrm{old}}}(\cdot \mid q)$.

In SPC-GRPO, instead of learning a separate value function, the advantage is estimated using an in-group baseline:

\begin{equation}
\bar{R}=\frac{1}{G}\sum_{i=1}^{G}R(o_i), \qquad
s_R=\sqrt{\frac{1}{G}\sum_{i=1}^{G}(R(o_i)-\bar{R})^2+\varepsilon},
\label{eq:group_stats}
\end{equation}
\begin{equation}
\hat{A}_i=\frac{R(o_i)-\bar{R}}{s_R}.
\label{eq:adv}
\end{equation}

The policy is updated by maximizing the following objective function, which promotes candidate solutions with higher relative physical quality while regularizing the policy update through a reference-policy KL constraint:

\begin{equation}
\resizebox{0.95\columnwidth}{!}{$\mathcal{J}(\theta)=
\mathbb{E}
\left[
\frac{1}{G}\sum_{i=1}^{G}
\frac{\pi_{\theta}(o_i \mid q)}{\pi_{\theta_{\mathrm{old}}}(o_i \mid q)}
\hat{A}_i
-\beta \,
\mathbb{D}_{\mathrm{KL}}
(\pi_{\theta}(\cdot \mid q)\|\pi_{\mathrm{ref}}(\cdot \mid q))
\right]$},
\label{eq:grpo_obj}
\end{equation}
where $\pi_{\mathrm{ref}}$ is the frozen SFT policy. This formulation encourages outputs with higher relative physical quality while preventing excessive deviation from the reference distribution.

\subsection{Multi-Granularity Hybrid Reward}

Central to our alignment strategy is the design of the reward function. Specifically, for a generated solution $o$, we formulate a Multi-Granularity Hybrid Reward (MGHR) as follows:

\begin{equation}
R(o)=
w_{\mathrm{fmt}} r_{\mathrm{fmt}}(o)+
w_{\mathrm{ast}} r_{\mathrm{ast}}(o)+
w_{\mathrm{exec}} r_{\mathrm{exec}}(o),
\label{eq:mghr}
\end{equation}
where $w_{fmt}$, $w_{ast}$, and $w_{exec}$ denote the weighting coefficients assigned to the respective components of the reward function. In this study, these weights are set to 0.05, 0.25, and 0.70, respectively. This distribution reflects the hierarchy of core engineering principles, where physical correctness constitutes the ultimate criterion, logical correctness serves as a necessary precondition, and formatting constraints fulfill a supplementary regulatory role.

\begin{table*}[htbp]
  \centering
  \caption{Main results on BMEval.}
  \label{tab:model_performance}
  \renewcommand{\arraystretch}{1.2}
  \setlength{\tabcolsep}{4pt}
  \small
  \begin{tabularx}{0.95\textwidth}{Xccccccc}
    \toprule
    \multirow{2}{*}{\textbf{Model Name}}
    & \multicolumn{2}{c}{\textbf{Executability}}
    & \multicolumn{4}{c}{\textbf{Engineering-Level Output Evaluation}}
    & \multirow{2}{*}{\textbf{\shortstack{Overall\\Average}}} \\
    \cmidrule(lr){2-3} \cmidrule(lr){4-7}
    & \textbf{Pass@1} & \textbf{Pass@5}
    & \textbf{Pass@5$_{\text{period}}$} & \textbf{Pass@5$_{\text{compliance}}$} & \textbf{Pass@5$_{\text{strict}}$} & \textbf{Average}
    & \\
    \midrule
    \textit{\textbf{DeepSeek-Coder-6.7B}} & 8.44{\scriptsize $\pm 2.4$} & 18.91{\scriptsize $\pm 2.9$} & 1.09{\scriptsize $\pm 1.1$} & 2.03{\scriptsize $\pm 2.7$} & 0.47{\scriptsize $\pm 1.3$} & 1.20 & 5.36 \\
    \textit{\textbf{Qwen2.5-Coder-7B-Instruct}} & 11.10{\scriptsize $\pm 1.1$} & 19.37{\scriptsize $\pm 2.2$} & 2.19{\scriptsize $\pm 2.2$} & 4.69{\scriptsize $\pm 2.6$} & 0.47{\scriptsize $\pm 0.5$} & 2.45 & 6.71 \\
    \textit{\textbf{CodeLlama-7B-Instruct}} & 7.65{\scriptsize $\pm 0.8$} & 17.66{\scriptsize $\pm 3.4$} & 2.19{\scriptsize $\pm 2.6$} & 4.22{\scriptsize $\pm 1.9$} & 0.31{\scriptsize $\pm 0.5$} & 2.24 & 5.71 \\
    \textit{\textbf{Opencoder-8B-Instruct}} & 9.85{\scriptsize $\pm 1.9$} & 20.47{\scriptsize $\pm 2.7$} & 2.81{\scriptsize $\pm 2.2$} & 3.44{\scriptsize $\pm 2.3$} & 1.56{\scriptsize $\pm 1.9$} & 2.60 & 6.79 \\
    \textit{\textbf{Seed-coder-8B}}$^{\mathrm{R}}$ & 11.88{\scriptsize $\pm 0.8$} & 22.19{\scriptsize $\pm 1.9$} & 2.03{\scriptsize $\pm 1.6$} & 3.44{\scriptsize $\pm 2.6$} & 0.62{\scriptsize $\pm 0.4$} & 2.03 & 7.03 \\
    \textit{\textbf{DeepSeek-V3.2}}$^{\mathrm{R}}$ & 7.81{\scriptsize $\pm 2.5$} & 11.72{\scriptsize $\pm 2.2$} & 3.28{\scriptsize $\pm 2.4$} & 2.34{\scriptsize $\pm 1.8$} & 0.78{\scriptsize $\pm 2.2$} & 2.14 & 4.68 \\
    \textit{\textbf{DeepSeek-R1}}$^{\mathrm{R}}$ & 5.16{\scriptsize $\pm 1.6$} & 10.31{\scriptsize $\pm 2.5$} & 1.25{\scriptsize $\pm 1.5$} & 2.66{\scriptsize $\pm 2.8$} & 0.16{\scriptsize $\pm 0.4$} & 1.35 & 3.48 \\
    \textit{\textbf{GLM-4.6}} & 2.35{\scriptsize $\pm 1.7$} & 9.06{\scriptsize $\pm 2.0$} & 1.87{\scriptsize $\pm 2.2$} & 2.19{\scriptsize $\pm 2.2$} & 0.00{\scriptsize $\pm 0.0$} & 1.35 & 2.80 \\
    \textit{\textbf{Qwen3-Coder-Plus}} & 12.97{\scriptsize $\pm 1.5$} & 28.91{\scriptsize $\pm 1.8$} & 2.03{\scriptsize $\pm 2.3$} & 5.94{\scriptsize $\pm 1.9$} & 0.31{\scriptsize $\pm 0.5$} & 2.76 & 8.82 \\
    \textit{\textbf{Qwen3-Coder-480B-A35B-Instruct}} & 12.97{\scriptsize $\pm 2.3$} & 24.84{\scriptsize $\pm 1.6$} & 3.44{\scriptsize $\pm 2.0$} & 5.16{\scriptsize $\pm 2.1$} & 1.25{\scriptsize $\pm 0.5$} & 3.28 & 8.49 \\
    \midrule
    \textit{\textbf{GPT-5}}$^{\mathrm{R}}$ & 32.50{\scriptsize $\pm 2.8$} & 59.06{\scriptsize $\pm 3.4$} & 22.97{\scriptsize $\pm 3.1$} & 32.03{\scriptsize $\pm 0.7$} & 12.19{\scriptsize $\pm 2.8$} & 22.40 & 30.19 \\
    \textit{\textbf{GPT-5-nano}} & 11.41{\scriptsize $\pm 1.8$} & 33.75{\scriptsize $\pm 2.0$} & 4.22{\scriptsize $\pm 1.1$} & 11.25{\scriptsize $\pm 1.9$} & 1.56{\scriptsize $\pm 0.7$} & 5.68 & 11.31 \\
    \textit{\textbf{Gemini-2.5-Flash}} & 41.41{\scriptsize $\pm 1.8$} & 75.16{\scriptsize $\pm 2.7$} & 7.03{\scriptsize $\pm 2.5$} & 17.50{\scriptsize $\pm 1.6$} & 4.69{\scriptsize $\pm 1.2$} & 9.74 & 25.92 \\
    \textit{\textbf{Gemini-3-pro-Preview}}$^{\mathrm{R}}$ & 26.72{\scriptsize $\pm 3.6$} & 54.22{\scriptsize $\pm 2.6$} & \textbf{32.03{\scriptsize $\pm 3.0$}} & 34.85{\scriptsize $\pm 2.8$} & 17.19{\scriptsize $\pm 3.5$} & 28.02 & 32.17 \\
    \textit{\textbf{Grok-Code-Fast}} & 19.53{\scriptsize $\pm 2.4$} & 42.97{\scriptsize $\pm 2.3$} & 13.91{\scriptsize $\pm 2.1$} & 19.22{\scriptsize $\pm 3.4$} & 9.38{\scriptsize $\pm 2.7$} & 14.17 & 19.86 \\
    \textit{\textbf{Claude-Sonnet-4.5}}$^{\mathbf{R}}$ & \textbf{52.34{\scriptsize $\pm 1.0$}} & \textbf{80.16{\scriptsize $\pm 2.8$}} & 21.25{\scriptsize $\pm 1.6$} & \textbf{48.75{\scriptsize $\pm 3.7$}} & \textbf{17.97{\scriptsize $\pm 1.4$}} & \textbf{29.32} & \textbf{41.63} \\
    \bottomrule
    \multicolumn{8}{p{0.95\textwidth}}{\footnotesize \textit{Note:} ``R'' denotes reasoning models. Boldface indicates the best result. Results are averaged over 5 perturbed evaluations of BMEval; $\pm$ denotes standard deviation. ``Average'' is the mean of Pass@5$_{\text{period}}$, Pass@5$_{\text{compliance}}$, Pass@5$_{\text{strict}}$. ``Overall Average'' is the mean of Pass@1, Pass@5, Pass@5$_{\text{period}}$, Pass@5$_{\text{compliance}}$, Pass@5$_{\text{strict}}$, and ``Average''.}
  \end{tabularx}
\end{table*}

\begin{table*}[htbp]
  \centering
  \caption{Performance comparison on AutoBM before and after applying RLA-SPC.}
  \label{tab:main_results}
  \renewcommand{\arraystretch}{1.2}
  \setlength{\tabcolsep}{4pt}
  \small
  \begin{tabular}{lccccccc}
    \toprule
    \multirow{2}{*}{\textbf{Model Name}}
    & \multicolumn{2}{c}{\textbf{Executability}}
    & \multicolumn{4}{c}{\textbf{Engineering-Level Output Evaluation}}
    & \multirow{2}{*}{\textbf{\shortstack{Overall\\Average}}} \\
    \cmidrule(lr){2-3} \cmidrule(lr){4-7}
    & \textbf{Pass@1} & \textbf{Pass@5}
    & \textbf{Pass@5$_{\text{period}}$} & \textbf{Pass@5$_{\text{compliance}}$} & \textbf{Pass@5$_{\text{strict}}$} & \textbf{Average}
    & \\
    \midrule
    \textbf{\textit{CodeLlama-7B-Instruct}} & 7.65{\scriptsize $\pm 0.8$} & 17.66{\scriptsize $\pm 3.4$} & 2.19{\scriptsize $\pm 2.6$} & 4.22{\scriptsize $\pm 1.9$} & 0.31{\scriptsize $\pm 0.5$} & 2.24 & 5.71 \\
    \quad + \textbf{\textit{RLA-SPC}} & 53.13{\scriptsize $\pm 4.3$} & 91.41{\scriptsize $\pm 1.5$} & 66.72{\scriptsize $\pm 5.8$} & 80.47{\scriptsize $\pm 5.3$} & 69.06{\scriptsize $\pm 3.3$} & 72.08 & 72.14 \\
    \midrule
    \textbf{\textit{Qwen2.5-Coder-7B-Instruct}} & 11.10{\scriptsize $\pm 1.1$} & 19.37{\scriptsize $\pm 2.2$} & 2.19{\scriptsize $\pm 2.2$} & 4.69{\scriptsize $\pm 2.6$} & 0.47{\scriptsize $\pm 0.5$} & 2.45 & 6.71 \\
    \quad + \textbf{\textit{RLA-SPC}} & 56.25{\scriptsize $\pm 6.6$} & 90.31{\scriptsize $\pm 4.5$} & 72.81{\scriptsize $\pm 2.5$} & 83.13{\scriptsize $\pm 4.3$} & 74.53{\scriptsize $\pm 4.4$} & 76.82 & 75.64 \\
    \midrule
    \textbf{\textit{Seed-Coder-8B-R}}$^{\mathrm{R}}$ & 11.88{\scriptsize $\pm 0.8$} & 22.19{\scriptsize $\pm 1.9$} & 2.03{\scriptsize $\pm 1.6$} & 3.44{\scriptsize $\pm 2.6$} & 0.62{\scriptsize $\pm 0.4$} & 2.03 & 7.03 \\
    \quad + \textbf{\textit{RLA-SPC}} & 62.19{\scriptsize $\pm 2.5$} & 96.72{\scriptsize $\pm 1.9$} & 73.75{\scriptsize $\pm 5.7$} & 92.19{\scriptsize $\pm 1.9$} & 75.00{\scriptsize $\pm 4.8$} & 80.31 & 80.03 \\
    \bottomrule
  \end{tabular}
  \vspace{2pt}
\end{table*}

\subsubsection{Format Compliance Reward}

The reward $r_{\mathrm{fmt}}$ is designed to enforce a strict output structure, ensuring compliance with Markdown code blocks and reasoning tags. It is formally defined as:

\begin{equation}
r_{\mathrm{fmt}}(o)=\mathbb{I}[\mathrm{FormatOK}(o)],
\label{eq:fmt}
\end{equation}
where $\mathrm{FormatOK}(\cdot)$ denotes a deterministic parser implemented via regular expressions.

\begin{algorithm}[t]
\caption{RLA-SPC}
\label{alg:rla-spc}
\KwIn{Instruction dataset $\mathcal{D}=\{(x^{(n)},y^{(n)})\}_{n=1}^N$; group size $G$; KL weight $\beta$; MGHR weights $(w_{\mathrm{fmt}},w_{\mathrm{ast}},w_{\mathrm{exec}})$.}
\KwOut{Aligned policy $\pi_{\theta}(\cdot\mid x)$.}

\BlankLine
\textbf{Stage I (SFT):} Initialize $\theta$ from a pretrained model\;
\While{not converged}{
Sample minibatch $\mathcal{B}\subset\mathcal{D}$\;
Update $\theta \leftarrow \theta - \eta \nabla_\theta \mathcal{L}_{\mathrm{SFT}}(\theta;\mathcal{B})$ using Eq.~\eqref{eq:sft}\;
}
Freeze $\pi_{\mathrm{ref}} \leftarrow \pi_{\theta}$\;

\BlankLine
\textbf{Stage II (SPC-GRPO):}
\While{not converged}{
Sample a batch of queries $\mathcal{Q}=\{q\}$ where $q\equiv x$\;
Set $\theta_{\mathrm{old}} \leftarrow \theta$\;

\ForEach{$q\in\mathcal{Q}$}{
Sample $G$ candidates $\{o_i\}_{i=1}^G \sim \pi_{\theta_{\mathrm{old}}}(\cdot\mid q)$\;

\ForEach{$o_i$}{
Compute format reward $r_{\mathrm{fmt}}(o_i)$ via Eq.~\eqref{eq:fmt}\;
Compute tier-wise coverage $c_k(o_i)$ via Eq.~\eqref{eq:tier_cov} and AST reward $r_{\mathrm{ast}}(o_i)$ via Eq.~\eqref{eq:ast_reward}\;
Execute $o_i$ in sandbox and compute $r_{\mathrm{exec}}(o_i)$ via Eqs.~\eqref{eq:exec_cases}--\eqref{eq:phy_reward}
(including progress reward Eq.~\eqref{eq:prog_reward} and period error Eq.~\eqref{eq:epsilon})\;
Aggregate $R(o_i)$ via Eq.~\eqref{eq:mghr}\;
}
Compute $\bar{R}, s_R$ via Eq.~\eqref{eq:group_stats} and advantages $\hat{A}_i$ via Eq.~\eqref{eq:adv}\;
Update $\theta \leftarrow \arg\max_\theta \mathcal{J}(\theta)$ using Eq.~\eqref{eq:grpo_obj}\;
}
}
\end{algorithm}

\subsubsection{Logical Completeness via Hierarchical AST}
Before execution, we assess whether the generated code is logically complete. Let $\mathrm{API}(o)$ be the set of APIs extracted from the AST of $o$. We construct three-tiered API sets: $\mathcal{T}_1$ for topology APIs, $\mathcal{T}_2$ for boundary and load APIs, and $\mathcal{T}_3$ for analysis and solver APIs.

Tier-wise coverage is defined as:

\begin{equation}
c_k(o)=\frac{|\mathrm{API}(o)\cap \mathcal{T}_k|}{|\mathcal{T}_k|}.
\label{eq:tier_cov}
\end{equation}

We further penalize undefined variables detected by static analysis. The logical reward is:

\begin{equation}
\tilde{r}_{\mathrm{ast}}(o)=
\sum_{k=1}^{3}\alpha_k c_k(o)
-\lambda N_{\mathrm{undef}}(o),
\qquad
r_{\mathrm{ast}}(o)=\mathrm{clip}(\tilde{r}_{\mathrm{ast}},0,1).
\label{eq:ast_reward}
\end{equation}

\subsubsection{Sandbox-Based Physical Consistency}
Finally, we evaluate physical correctness by executing the code in an OpenSees sandbox:

\begin{equation}
r_{exec} = \begin{cases} 
\eta_{env}, & \textit{if env. error or timeout} \\
R_{prog}(l_{err}, L_{tot}), & \textit{if runtime logic error} \\
R_{phy}(\epsilon), & \textit{if success with physical metrics}
\end{cases}
\label{eq:exec_cases}
\end{equation}

For runtime failures, we use execution progress:

\begin{equation}
R_{\mathrm{prog}}=
\alpha_{\mathrm{base}}+
\beta_{\mathrm{step}}\frac{l_{\mathrm{err}}}{L_{\mathrm{tot}}}.
\label{eq:prog_reward}
\end{equation}
where $l_{\mathrm{err}}$ represents the specific line number of the reported error as captured by the interpreter traceback, and $L_{\mathrm{tot}}$ denotes the total number of code lines. The parameter $\alpha_{\mathrm{base}}$ signifies the base attempt reward, which is utilized to incentivize the successful initiation of the execution stage. Furthermore, $\beta_{\mathrm{step}}$ is defined as the progress scaling factor, employed to modulate the contribution of code completion to the overall score.

For successful executions, physical consistency is evaluated using the relative error of the structural fundamental period.

\begin{equation}
\epsilon=\frac{|T_{\mathrm{pred}}-T_{\mathrm{gt}}|}{T_{\mathrm{gt}}},
\label{eq:epsilon}
\end{equation}

\begin{equation}
R_{\mathrm{phy}}(\epsilon)=
\begin{cases}
1.00, & \epsilon \le 0.10,\\
0.90, & 0.10<\epsilon \le 0.20,\\
0.80, & 0.20<\epsilon \le 0.40,\\
0.70, & \text{otherwise}.
\end{cases}
\label{eq:phy_reward}
\end{equation}

This hierarchical reward design explicitly reflects the tolerance-based decision paradigm commonly adopted in structural engineering practice, where approximate physical consistency is often acceptable within specified error bounds, while larger deviations are progressively penalized.

\begin{figure*}[htbp]
  \centering
  \includegraphics[width=\textwidth]{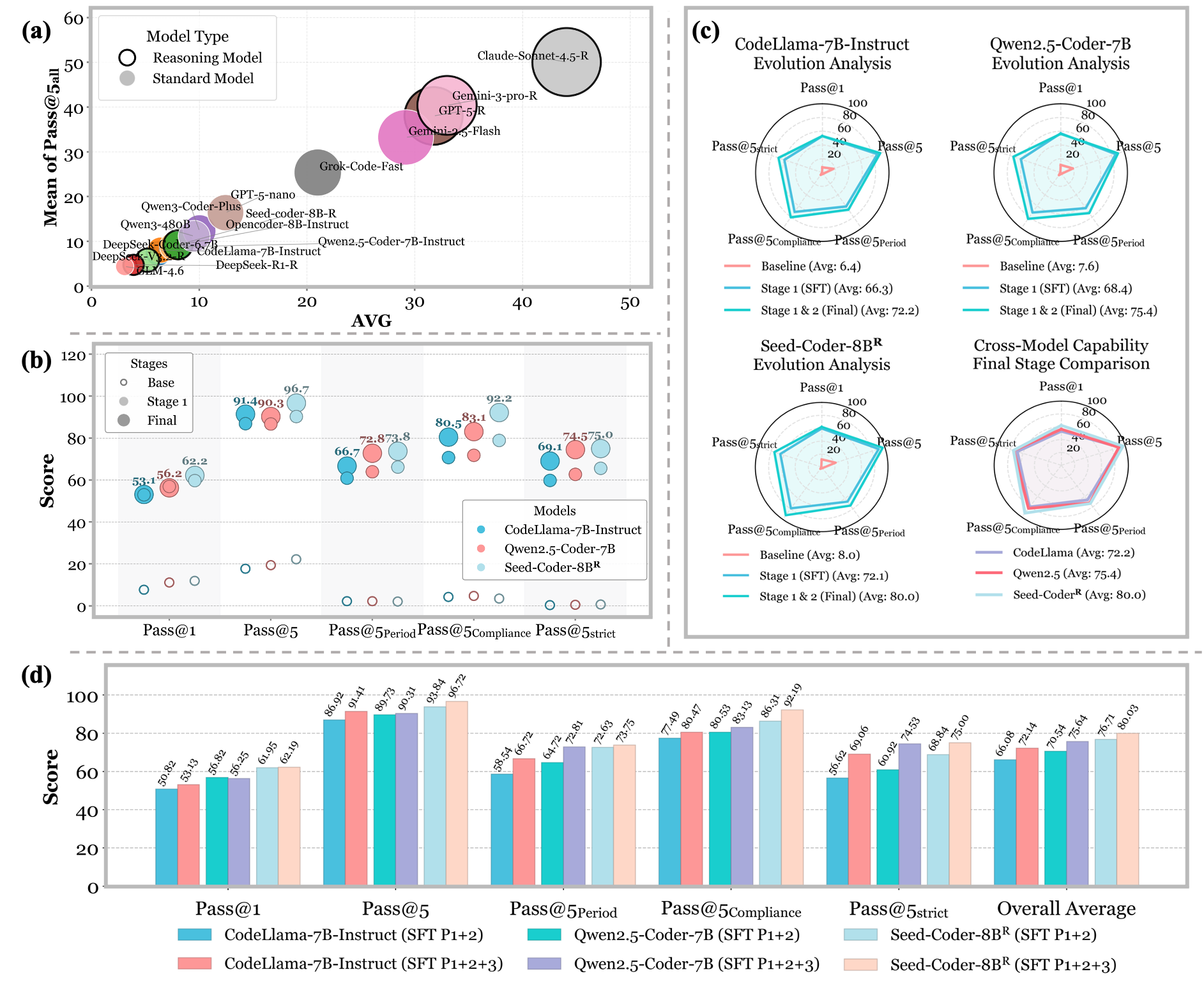} 
  \caption{The results of the report generation tasks. (a) LLM performance on the BMEval benchmark. (b) \& (c) Evolution of model performance across training stages. (d) Impact of data composition on baseline models}
  \label{fig:results of the report}
\end{figure*}

\begin{figure*}[!htb]
  \centering
  \includegraphics[width=\textwidth]{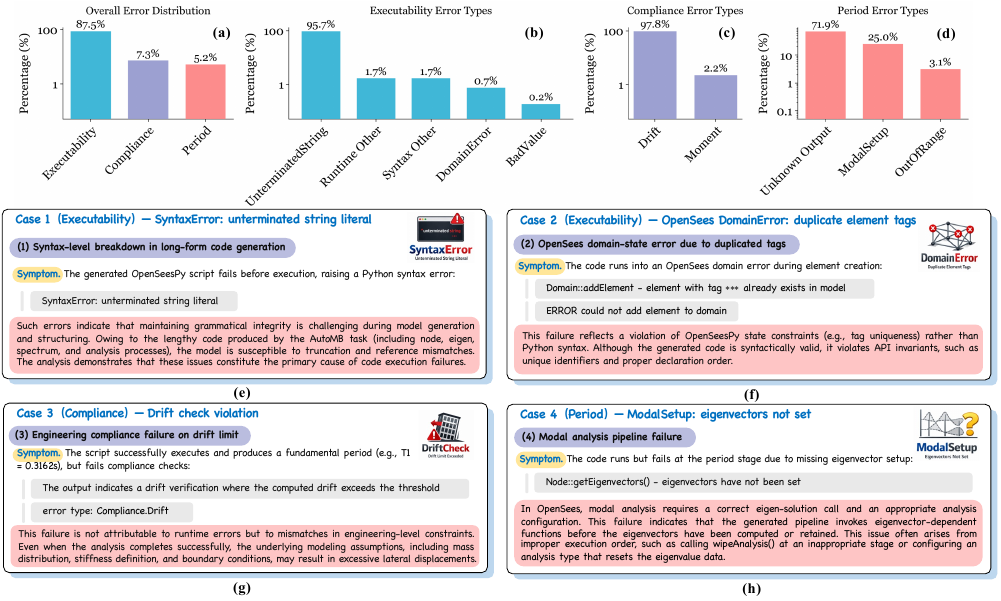} 
  \caption{The Limitations of AI-Generated AutoBM Task. An analysis based on 640 sets of modeling code generated by Gemini-2.5-Flash. Subfigures (a–d) present the distribution of error categories, (e–h) provide detailed analyses of individual cases.}
  \label{Gemini-2.5-Flash}
\end{figure*}

\section{Experiments and Analysis}
\subsection{Experimental Setup}
To evaluate LLM performance on the AutoBM task, a set of representative state-of-the-art open-source and commercial models was assessed using the BMEval benchmark, which consists of 128 automated modeling tasks designed to cover diverse structural configurations and seismic conditions. All models were evaluated under a unified protocol, and the main results are summarized in Table \ref{tab:model_performance} and Figure \ref{fig:results of the report}(a).

\subsection{Limitations of LLMs for the AutoBM Task}
Although leading models such as $\text{Claude-Sonnet-4.5}^{\mathrm{R}}$ and $\text{GPT-5}^{\mathrm{R}}$ achieve high code execution rates, their joint evaluation scores remain low, indicating significant limitations in practical engineering applicability under realistic design constraints.

\textbf{Grammatical and physical divide.} Models exhibit a systematic mismatch between syntactic correctness and physical validity. For instance, Gemini-2.5-Flash attains a high $Pass@5$ score (75.16\%) but an extremely low $Pass@5_{strict}$ score (4.69\%), suggesting that physical parameters are treated as statistical symbols rather than physically constrained quantities. During the inference process, these parameters are manipulated as discrete symbols governed by learned distributions, rather than as real-valued variables constrained by fundamental physical principles (non-negativity, continuity). Consequently, models frequently generate simulations that are syntactically valid but physically implausible under realistic design constraints.

\textbf{Spatial-topological reasoning.} Performance on the AutoBM task remains limited. Gemini-3-Pro-Preview achieves the highest structural prediction accuracy ($Pass@5 = 32.03\%$), while other leading models perform moderately, and most open-source models score below 3\%. These results indicate persistent difficulties in maintaining consistent geometric representations, with no clear scaling trend with model size.

\textbf{Engineering self-verification.} Closed-loop reasoning remains weak across models. Although $\text{Claude-Sonnet-4.5}^{\mathrm{R}}$ achieves relatively better compliance ($Pass@5_{compliance} = 48.75\%$), most models fail to perform specification-driven verification, even under explicit constraint prompts. Notably, even when explicit constraints to “meet the seismic design requirements” are incorporated into the prompt, these models often remain at the level of surface text generation and fail to activate the conditional branching logic required for genuine design verification and parameter adjustment.

To further investigate the training dynamics, Figures \ref{fig:results of the report}(b–d) and Table \ref{tab:main_results} illustrate the performance evolution and the impact of data composition. The analysis indicates that the second stage of training is critical for improving compliance-related metrics (see Figures \ref{fig:results of the report}(b) and \ref{fig:results of the report}(c)), and that data matching plays a pivotal role in determining training effectiveness (see Figure \ref{fig:results of the report}(d)). A comprehensive analysis of the training stages and the ablation studies presented in Figures \ref{fig:results of the report}(b–d) is provided in Appendix \ref{Supplementary Ablation Studies}.

\subsection{Case Study: Failure Mode Analysis in AutoBM}
To further investigate the limitations of AutoBM, a systematic code inspection of the LLM-generated code was conducted. As shown in Figure \ref{Gemini-2.5-Flash}(a), the observed failure modes were primarily categorized into three types: executability, compliance, and period.



\textbf{Executability} (Figure \ref{Gemini-2.5-Flash}(b)): \texttt{SyntaxError: unterminated string literal} accounted for 95.7\% of executability failures. This finding suggests that preserving syntactic integrity remains a major challenge for AutoBM when generating long scripts, particularly those with extensive node definitions and multi-step analysis pipelines.

\textbf{Compliance} (Figure \ref{Gemini-2.5-Flash}(c)): \texttt{Compliance.Drift} was the predominant failure mode. This outcome implies that even when execution succeeds and modal results are produced, the synthesized structural models frequently violate engineering constraints due to excessive or unacceptable lateral drift.

\textbf{Period} (Figure \ref{Gemini-2.5-Flash}(d)): Failures were primarily attributed to period. \texttt{Unknown}, followed by \texttt{ModalSetup} errors. These patterns indicate persistent difficulty in reliably extracting and validating modal periods, as well as in maintaining correct eigen-solution configurations required by the analysis workflow.

Figures \ref{Gemini-2.5-Flash}(e–h) present four representative failure cases. Specifically, these failures span multiple layers, ranging from syntactic integrity to simulator state constraints to engineering compliance. Overall, the strong dominance of a small set of failure modes provides a clear basis for developing targeted mitigation strategies.

\section{Limitations and Ethical Considerations}
Despite the promising performance of AutoBM, several limitations remain. The current framework is primarily designed for elastic structural and simplified 2D reinforced concrete frames, whereas nonlinear analysis and 3D interactions are not yet incorporated. In addition, the framework has been developed mainly under Chinese seismic design specifications, and its generalizability to other international design codes has not been systematically validated. Although improvements in executability and physical consistency have been achieved, hidden engineering defects may still arise under complex conditions. Therefore, AutoBM is intended to assist engineering modeling workflows rather than replace professional structural design, expert review, or regulatory verification in safety-critical applications.

\section{Conclusion}
In this paper, we propose a domain-specific dataset, CivilInstruct, and a two-stage fine-tuning strategy guided by structural and physical constraints are proposed. In addition, a physics-aware evaluation benchmark, BMEval, is developed to provide engineering-relevant performance assessment. Results show that supervised instruction fine-tuning significantly improves baseline modeling ability and code executability. Building on this, physically constrained reinforcement learning further enhances physical consistency and specification compliance. Ablation studies confirm the complementary roles of the two stages, with SFT enabling fundamental modeling competence and reinforcement learning reducing physics illusion.

\section*{Acknowledgments}
The authors gratefully acknowledge the support from the National Natural Science Foundation of China (Grant Nos. U24A20177 and 52408553) and the International Collaboration Program of Sichuan Province (Grant No. 2025YFHZ0132). Y.Q. acknowledges the support of the China Scholarship Council program (Project ID: 202506240041).

\section*{GenAI Disclosure}
The authors used ChatGPT-5.2 to refine the language and improve the grammatical quality of the manuscript. In addition, Nano Banana was used to assist in generating the visual elements included in the illustrations for Figures 1, 2, and 4. After using this tool, the authors carefully reviewed and edited the content as necessary and assume full responsibility for the material presented in the published paper.

\bibliographystyle{ACM-Reference-Format}
\balance
\bibliography{sample-base}

\appendix

\section*{Appendix}
\setcounter{section}{0}

\section{BMBench Implementation Details}
\label{sec:BMBench}
\subsection{CivilInstruct Dataset Construction Details}
\label{sec:bmbench_impl}

\textbf{Part 1: Fine-Grained Supervised Data.} A supervised dataset comprising 3,881 instruction-code pairs was constructed through the systematic parsing of official OpenSeesPy documentation in conjunction with LLM synthesis. Designed to facilitate instruction adherence and automated structural modeling, each data instance comprises two core components: task-oriented natural language instructions derived from standard usage protocols and executable code templates adhering to strict documentation specifications. To enhance model interpretability and eliminate ambiguity during the learning process, detailed line-by-line annotations were incorporated into the code. These annotations provide engineering-level semantic interpretations of key variables and parameters. Comprehensive coverage of core functional modules is provided, encompassing structural modeling, numerical analysis, reliability analysis, and pre- and post-processing. In contrast to general-purpose codebases, which frequently lack domain semantics or exhibit non-standard API usage, CivilInstruct rigorously ensures the clarity of variable semantics and the consistency of documentation specifications.

\begin{table}[htbp]
\centering
\caption{Engineering Parameter Ranges for Random Structural Sample Generation}
\label{tab:parameter_spec}
\renewcommand{\arraystretch}{1.2}
\begin{tabularx}{\columnwidth}{>{\raggedright\arraybackslash}X
                                  >{\raggedright\arraybackslash}X}
\toprule
\textbf{Parameter} & \textbf{Value Range / Set} \\
\midrule
Building Function
& \{Office, Residential, Mall, Hospital, School, Factory\} \\

Number of Stories
& Integer $\in [3, 7]$ \\

Story Height
& Uniform $\sim [3.0, 4.0]$ m \\

Total Height
& $H < 23.0$ m \\

Plan Dimensions
& Uniform $\sim [40, 100]$ m, subject to $W \leq L$ \\

Seismic Intensity
& \{7, 7.5, 8, 8.5, 9\} \\

Peak Ground Acceleration
& \{0.10, 0.15, 0.20, 0.30, 0.40\} g \\

Seismic Group
& \{1, 2, 3\} \\

Site Class
& \{$I_0$, $I_1$, II, III, IV\} \\
\bottomrule
\end{tabularx}
\end{table}

\textbf{Part 2: Parameterized Generated Long Code Data.} Approximately 3,100 structural instances were constructed via the stochastic sampling of key architectural and seismic parameters within predefined valid engineering domains. These sampling parameters include building function, story count, and story heights (see Table \ref{tab:parameter_spec} for complete parameter ranges and discrete value sets). While the selected parameter domains align with prevalent seismic design practices and national codes (e.g., JGJ/T 415-2017 \cite{zhongming2009ministry}), excessively rigid code-specific constraints were avoided to ensure the generalizability of data-driven learning. For each instance, the LLM first synthesized structural design solutions that conform to engineering principles. Subsequently, building attributes and design solutions were integrated into a specific instruction template (see Appendix \ref{sec:BMBench}), which guided the model in generating complete, task-specific code. To enhance the diversity of instruction distributions, 1,000 instruction variants were derived from the original template, adopting the approach of Xu et al. \cite{xu2024wizardlm}. The resulting programs typically comprise approximately 600 lines of executable code, encompassing the entire workflow from structural modeling to analysis and validation. To ensure rigorous data quality control, automated validation of all generated code was performed, with quantitative scores assigned based on code quality and task completion. Ultimately, only high-quality samples were retained for the second stage of reinforcement learning training. At the same time, data about execution failures were segregated for use in constructing the Part 3 data component.

\textbf{Part 3: Execution Error-Oriented Debugging of CoT Data.} A specialized Bug-CoT dataset was constructed through the systematic analysis of execution failures identified in Part 2. Detailed debugging information was captured for these failure samples, encompassing error logs, return codes, runtime environments, and contextual metadata. A distinguishing characteristic of this dataset is its bug-driven generation mechanism. Unlike synthetic defects, these errors arise from complex, coupled interactions between the OpenSeesPy API, the numerical solver, and the modelling logic. Consequently, they are representative of authentic challenges encountered in actual engineering analyses. For each documented error, the LLM was employed to generate a structured debugging CoT, comprising a problem description, root cause analysis, resolution strategy, and prevention guidelines. Furthermore, to enhance data coverage and inference robustness, each original error was expanded into multiple semantically related yet distinct variants. Ultimately, all inference pathways were converted into a question-and-answer format, providing fine-grained supervisory signals to train the model in the systematic debugging of extended scientific computing sequences.

\textbf{Part 4: Physics-Informed Expert Data.} Based on the extensive codebases generated in Part 2, a high-quality expert dataset was curated for the subsequent reinforcement learning stage. Specifically, through the implementation of automated execution and multidimensional quantitative scoring, which encompasses code executability and compliance with design specifications, 512 sample sets achieving scores exceeding 90 out of 100 were isolated from the pool of executable code. These samples not only demonstrate runtime robustness but also adhere to stringent engineering constraints. To provide robust and physically interpretable training signals for reinforcement learning, the structural fundamental period, a critical metric in dynamics, was extracted to serve as the ground truth. As a pivotal parameter in seismic design, this indicator possesses distinct physical significance. Consequently, it serves as an effective supervisory signal, guiding the model toward acquiring a structural logic that aligns with engineering rationality.

\subsection{BMEval Construction Details}
\label{sec:BMEval Construction Details}

\textbf{Structural Cycle Consistency Assessment ($Pass@k_{period}$)}. Because program executability alone is insufficient to satisfy stringent engineering requirements, the numerical rationality of the generated outputs was further evaluated. This study focuses on the first-order natural vibration period of the structure, denoted as $T_1$, and extracts the predicted values from program outputs using a deterministic, rule-based matching method. If the relative error between the predicted value and the ground-truth value satisfies the following criterion:

\begin{equation}
\frac{|T_1^{Pred} - T_1^{gt}|}{T_{gt}} \le \epsilon
\end{equation}

Here, $T_{gt}$ denotes the reference ground-truth, and the allowable relative error threshold adopted in this experiment is set to 0.30. The resulting $Pass@k_{period}$ metric, calculated according to this criterion, is intended to quantify the model's engineering-level numerical accuracy.

\textbf{Engineering Compliance Assessment ($Pass@k_{compliance}$)}. In addition to numerical accuracy, engineering analysis results must provide explicit conclusions regarding compliance with specifications or safety verification. To address this requirement, engineering compliance evaluation indicators were introduced to assess whether the program output clearly articulates a conclusion that satisfies design verification criteria. Methodologically, a conservative keyword-matching strategy was adopted, in which affirmative terms (e.g., “pass”, “satisfy”) were identified, while negative semantic processing mechanisms were incorporated to reduce the risk of misclassification. If a program, following successful execution, produces a verification conclusion that conforms to the relevant engineering specifications, it is considered to have passed the compliance assessment.

\begin{figure}[htbp] 
 \centering
 \includegraphics[width=\linewidth]{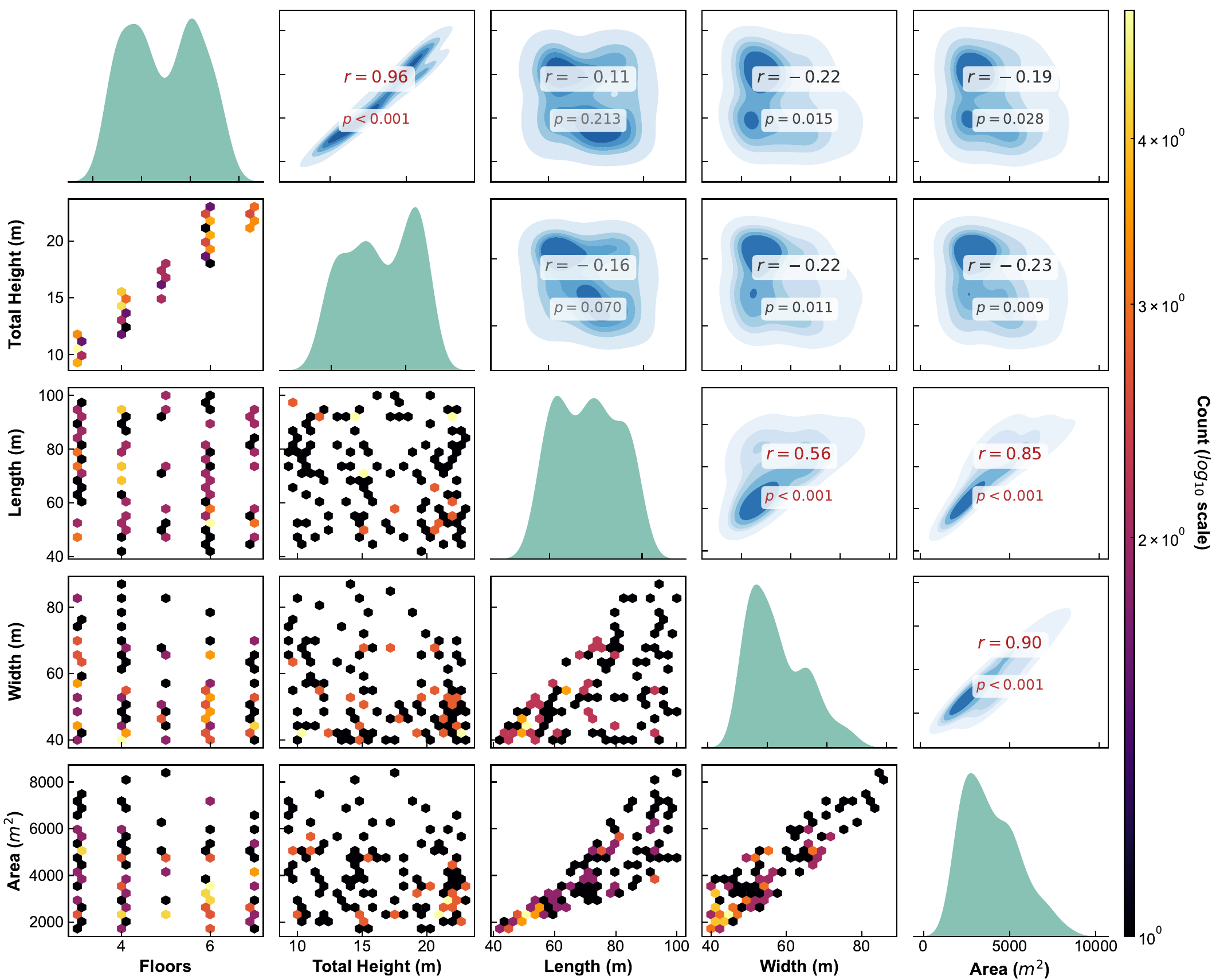} 
 \caption{Building Attributes Correlation Matrix}
 \label{Correlation Matrix}
\end{figure}


\textbf{Joint Evaluation Indicator ($Pass@k_{strict}$)}. To comprehensively assess the AutoBM capabilities of the model under realistic engineering application scenarios, a more stringent composite evaluation criterion was defined. A generated code sample is considered strictly correct only if it simultaneously satisfies the following three necessary conditions: (1) the program executes successfully without exceptions; (2) the extracted structural period prediction falls within the prescribed allowable error range; and (3) the output includes an explicit conclusion confirming compliance with relevant design specifications. The $Pass@k_{strict}$ indicator, calculated according to this criterion, represents the most rigorous comprehensive performance metric employed in this study and most effectively reflects the model’s practical value in real-world engineering applications.

In Figure \ref{Correlation Matrix}, the correlation matrix of building attributes is presented, demonstrating the rationality and diversity of the BMEval distribution.



\section{Prompt used in AutoBM}
\label{sec:Prompt used in AutoBM}

\subsection{Prompt Template for Structural Design and Modeling Code Generation}
To generate analysis-ready reinforced concrete (RC) frame models in a consistent and code-compliant manner, we adopt a rule-constrained, prompt-based design workflow. Domain-specific knowledge from seismic engineering and relevant Chinese design codes (GB50011-2010, GB50010-2010, and JGJ3-2010) is embedded as explicit constraints, requiring the model to output only final, validated structural parameters formatted for direct implementation in OpenSeesPy. The workflow enforces capacity design principles, gravity load limits, seismic detailing requirements, and interstory drift constraints. An implicit verification loop is incorporated, whereby component dimensions and reinforcement configurations are iteratively adjusted until all strength and displacement criteria are satisfied.

\section{Supplementary Ablation Studies}
\label{Supplementary Ablation Studies}
\subsection{RLA-SPC Can Improve Model Performance on AutoBM Task}
This section systematically demonstrates the effectiveness of the proposed two-stage fine-tuning strategy with structural physical constraints for the AutoBM task through a series of ablation experiments. In addition, it provides an in-depth analysis of the differentiated contributions of SFT and reward-based optimization to the development of model capabilities. Three experimental configurations were evaluated across three representative baseline models (CodeLlama-7B-Instruct, Qwen2.5-Coder-7B, and $\text{Seed-Coder-8B}^{\mathrm{R}}$): the original baseline model, a model trained using SFT only (stage 1, incorporating data Parts 1-3), and a physically aligned two-stage model that integrates GRPO (stage 1+2), augmented with data Part 4 and physics-constrained reward signals.

The experimental results (Table \ref{tab:main_results}, Figure \ref{fig:results of the report}(b-c)) indicate that SFT plays a decisive role in establishing the model’s foundational modeling capability. Compared with the baseline, the code execution rate (Executability) of each model increases substantially after stage 1 training, demonstrating that the models effectively acquire the syntactic structure and common programming patterns of OpenSeesPy. Although the code generated by SFT-trained models is generally executable, it is often accompanied by “physical illusion” phenomena that violate fundamental mechanical principles or engineering design specifications. With the introduction of GRPO, the models exhibit stable and significant improvements not only in code executability but also in stringent performance metrics (e.g., $Pass@5_{strict}$). These results suggest that GRPO effectively enforces physical consistency and specification compliance through explicit reward-based guidance. A comparative analysis further reveals the complementary roles of the two stages: stage 1 establishes basic syntactic executability, whereas stage 2 achieves semantic alignment with physical laws and engineering specifications, thereby substantially reducing the generation of physically implausible outputs.

\subsection{Execution Error-Oriented Debugging Chain of Thought Data Ablation Study}
This section aims to systematically validate the effectiveness of the CivilInstruct data construction strategy, with particular emphasis on assessing the critical role of the execution-error-driven Bug Fix dataset (Part 3) in enhancing code robustness and engineering reliability. To this end, two SFT data construction schemes were designed for comparison, while maintaining strict consistency in the model architecture and learning strategies. The first scheme uses only Parts 1 and 2 as positive samples for SFT. The second scheme enhances this setup by incorporating Part 3 (Bug-CoT data), allowing the model not only to learn correct coding paradigms during fine-tuning but also to be exposed to real execution failure cases, along with their systematic error analyses and repair trajectories. Experimental results (Figure \ref{fig:results of the report}(d)) demonstrate consistent and substantial performance improvements across all three baseline models (CodeLlama-7B-Instruct, Qwen2.5-Coder-7B, and $\text{Seed-Coder-8B}^{\mathrm{R}}$). Specifically, after introducing the Bug-CoT dataset, the $Pass@1$ score of CodeLlama-7B-Instruct increased from 50.82\% to 53.13\%, while its $Pass@5_{strict}$ score improved markedly from 56.62\% to 69.06\%. The Qwen2.5-Coder-7B and $\text{Seed-Coder-8B}^{\mathrm{R}}$ models exhibit similar gains, with $Pass@5_{strict}$ increases of 13.61\% and 6.16\%, respectively.

Experimental results indicate that incorporating Part 3 not only enhances code executability but also significantly increases the success rate under stringent engineering constraints. Notably, this improvement does not arise from superficial normalization of output formats, but rather from the model’s deeper assimilation of real failure patterns. By learning to recognize low-level syntactic errors, misuse of application programming interfaces, and the underlying causes of structurally unstable configurations, the model can proactively avoid fragile modeling decisions during the generation process, thereby producing more robust candidate solutions. This enriched and higher-quality solution space provides a strong foundation for the subsequent GRPO stage, enabling the reinforcement learning procedure to identify optimal trade-offs between physical constraints and regulatory objectives more efficiently.

\end{document}